\documentclass[prd,groupedaddress, showpacs, twocolumn, amsmath,amssymb,nofootinbib]{revtex4}

\usepackage{graphicx}
\usepackage{dcolumn}
\usepackage{bm}
\usepackage{amsmath,graphics,color}
\usepackage{color,psfrag}
\usepackage{mathrsfs}

\newcommand{\nn}{\nonumber}
\newcommand{\be}{\begin{equation}}
\newcommand{\ee}{\end{equation}}
\newcommand{\bea}{\begin{eqnarray}}
\newcommand{\eea}{\end{eqnarray}}

\begin{document}

\title{The dynamics of generalized Palatini Theories of Gravity}

\author{Vincenzo Vitagliano$^{1}$}
\email{vitaglia@sissa.it}
\author{Thomas P. Sotiriou$^{2}$}
\email{T.Sotiriou@damtp.cam.ac.uk}
\author{Stefano Liberati$^{1}$}
\email{liberati@sissa.it}
\affiliation{${}^1$SISSA-International School for Advanced Studies,  Via Bonomea 265, 34136 Trieste, Italy and INFN sezione di Trieste, via Valerio 2, 34127 Trieste, Italy}
\affiliation{${}^2$Department of Applied Mathematics and Theoretical Physics, Center for Mathematical Sciences, University of Cambridge, Wilberforce Road, Cambridge, CB3 0WA, UK}

\begin{abstract}
It is known that in $f(R)$ theories of gravity with an independent connection which can be both non-metric and non symmetric, this connection can always be algebraically eliminated in favour of the metric and the matter fields, so long as it is not coupled to the matter explicitly. We show here that this is a special characteristic of $f(R)$ actions, and it is not true for actions that include other curvature invariants.  This contradicts some recent claims in the literature. We clarify the reasons of this contradiction.

\end{abstract}
\maketitle


%


%



\section{Introduction}

Einstein's equations can be derived by varying the Einstein--Hilbert action with respect to the metric. They can also be derived by what is formally the same action, by assuming that the connection is independent of the metric and performing independent variations with respect to the metric and the connection. This is called a Palatini variation and it can be found in some textbooks, see for example Ref.~\cite{grav}. Note that in the Palatini variation the independent connection is assumed to not enter the matter action.

Even though both standard metric and Palatini variations of (what is formally) the Einstein--Hilbert action lead to equivalent systems of field equations, this is not the case for more general actions. A typical example of actions that have been widely studied with both variational principles are $f(R)$ actions, see Refs.~\cite{sfreview,phd,Sotiriou:2008ve,Nojiri:2006ri,DeFelice:2010aj} for reviews. Indeed there is by now a long literature on $f(R)$ theories with an independent, symmetric connection which does not couple to the matter, dubbed Palatini $f(R)$ theories of gravity \cite{palgen,fer}. 

Even though these theories are not equivalent to the theory corresponding to the same action obtained with simple metric variation, they are nevertheless still metric theories according to the Thorne-Will  definition \cite{will}.\footnote{Quoting Thorne and Will \cite{will}, a metric theory is a theory that satisfies the metric postulates, {\em i.e.} a theory for which ``(i) gravity is associated, at least in part, to a symmetric tensor, the \textit{metric} and (ii) the response of matter and fields to gravity is described by $\nabla_{\mu}T^{\mu\nu}=0$, where $\nabla_{\mu}$ is the divergence with respect to the metric and $T^{\mu\nu}$ is the stress-energy tensor for all matter and nongravitational fields.''} In fact, the independent connection in Palatini $f(R)$ gravity does not actually carry any dynamics. It is really an auxiliary field that can be eliminated in favour of the metric and the matter fields \cite{koivisto,sot2,Sotiriou:2007zu}. This result has recently been generalized to  $f(R)$ theories with non-symmetric connections, {\em i.e.}~theories that allow for torsion \cite{Sotiriou:2009xt}. Additionally, Palatini $f(R)$ gravity has been shown to be dynamically equivalent to Brans--Dicke theory with Brans--Dicke parameter $\omega_0=-3/2$ \cite{flanagan,olmonewt,sot1} irrespectively of how general the connection is allowed to be \cite{Sotiriou:2009xt}. This is a particular theory within the Brans--Dicke class in which the scalar does not carry any dynamics and can be algebraically eliminated in favour of the matter fields.

The fact that in Palatini $f(R)$ gravity the independent connection is non-dynamical can be viewed as a blessing at first: no extra degrees of freedom are introduced with respect to general relativity, so one need not worry about pathologies usually associated with such degrees of freedom (ghost modes, instabilities etc.) or conflicts with current experimental bounds on their existence. However, one soon realizes that having a theory with second order dynamics and still different from general relativity actually requires a drastic departure from the latter. Indeed, a number of viability issues plague generic models of Palatini $f(R)$ gravity, and all of these shortcomings have their origin at the peculiar differential structure of the theory \cite{Barausse:2007pn}. 

Palatini $f(R)$ gravity models with infrared corrections with respect to general relativity have been shown to be in conflict with the standard model of particle physics \cite{flanagan,padilla} and to violate solar system tests as their post-Newtonian metric  has an algebraic dependence on the matter fields \cite{olmonewt,sotnewt}. Singularities have been shown to arise on the surface of well known spherically symmetric matter configuration \cite{Barausse:2007pn}, which render the theory at best incomplete and provide a very strong viability criterion. This criterion is almost independent of the functional form of the Lagrangian, the only exception being Lagrangians with corrections which become important only in the far ultraviolet (as in this case the singularities manifest at scales where non-classical effects take over) \cite{Olmo:2008pv} . 

On the other hand, $f(R)$ actions are a special class and there is no reason for one to restrict to those. In fact from an effective field theory point of view such a restriction can be considered a severe fine tuning. It is, therefore, interesting to consider more general actions. Then a question naturally arises: will these more general actions share the property of Palatini $f(R)$ of actually having a non-dynamical connection? Or will at least some of the degrees of freedom hiding in the connection be excited? This is what we would like to address here. The answer is  ultimately related to whether such more general theories would suffer by the same shortcomings as Palatini $f(R)$ gravity, which, as mentioned, can be traced back to the presence of the non-dynamical connection. 

Generalized Palatini theories of gravity have been considered to some extent in the literature. In Ref.~\cite{Allemandi:2004wn} the cosmology of Lagrangians of the form $f(R^{(\mu\nu)}R_{(\mu\nu)})$  was studied, parentheses indicating symmetrization.  In Ref.~\cite{Li:2007xw} the focus was on theories of the form $R+f(R^{(\mu\nu)}R_{(\mu\nu)})$.  Finally, in Refs.~\cite{Olmo:2009xy,olmo2,new} Lagrangians of the more general form $f(R,R^{\mu\nu}R_{\mu\nu})$ were studied. In fact, in Ref.~\cite{Olmo:2009xy} the very question that we are posing here was considered and it was claimed that the connection can indeed be eliminated. We argue that this claim in not correct, at least unless one imposes extra {\em a priori} restrictions on the connection or the action. 

The rest of the paper is organized as follows. In the next section we illustrate briefly how the connection can be algebraically eliminated in the case of $f(R)$ theories. This will serve as a brief review of the results in the literature. In section \ref{main} we move on to consider more general actions and we argue that the connection cannot be eliminated for generic actions. We discuss some special cases that constitute exceptions and we show that they do not include the action considered in Refs.~\cite{Olmo:2009xy,olmo2,new}, contrary to what was claimed there. We also give an easy but characteristic example of a generalized Palatini theory with extra degrees of freedom with respect to general relativity. Section \ref{conc} contains our conclusions.

Before going further it is worth emphasizing that throughout this paper we are considering theories in which the independent connection does not enter the matter action, {\em i.e.}~it does not couple to the matter fields. One can clearly question if this is the most sensible choice, and in fact it would be very reasonable to allow for the independent connection to define the covariant derivative and, therefore, couple to (at least) some matter fields. $f(R)$ theories of this type, dubbed metric-affine $f(R)$ theories of gravity, have been introduced in \cite{sotlib}. We will not consider them or their generalizations here, as they are in fact the subject of a separate publication \cite{metaffdyn}.

\section{$f(R)$ actions as an example}
\label{fofr}

We start by briefly reviewing how the independent connection can be eliminated in Palatini $f(R)$ gravity. For simplicity we restrict ourselves to a symmetric connection, even though the results can be generalized to a non symmetric one. We refer the reader to Ref.~\cite{Sotiriou:2009xt} for details.

Consider the action
\be
\label{faction}
S=\frac{1}{l_p^2} \int dx^4  \sqrt{-g}f({\cal R})+S_M(\psi,g_{\mu\nu}),
\ee
where $g$ is the determinant of the metric $g_{\mu\nu}$, ${\cal R}_{\mu\nu}$ is the Ricci tensor of the independent connection, ${\cal R}=g^{\mu\nu}{\cal R}_{\mu\nu}$, $S_M$ is the matter action, $\psi$ collectively denotes the matter fields (note that the connection does not enter the matter action) and $l_p$ has dimensions of a length. 
Varying the action independently with respect to the metric and the connection gives the following set of field equations, after some manipulations:
\bea
\label{field1}
&&f'({\cal R}) {\cal R}_{(\mu\nu)}-\frac{1}{2}f({\cal R})g_{\mu\nu}=\kappa T_{\mu\nu},\\
\label{field22}
&&\nabla_\lambda\left(\sqrt{-g}f'({\cal R})g^{\mu\nu}\right)=0,\eea
where $\nabla_\mu$ is the covariant derivative defined with the independent connection, a prime denotes differentiation with respect to the argument,
\be
T_{\mu\nu}=-\frac{1}{8\pi \sqrt{-g}}\frac{\delta S_M}{\delta g^{\mu\nu}},
\ee
and $\kappa=8\pi\, l_p^2$.
The right-hand side of eq.~(\ref{field22}) vanishes thanks to our assumption that the matter action is independent of the connection. Details of the variation can be found in section 4.1 of Ref.~\cite{sotlib}.

Eq.~(\ref{field22}) can be solved for the connection to give
\bea
\label{gammagmn}
\Gamma^\lambda_{\phantom{a}\mu\nu}&=&\left\{^\lambda_{\phantom{a}\mu\nu}\right\}+\frac{1}{2f'}\Big[2\partial_{(\mu} f'\delta_{\nu)}^\lambda-g^{\lambda\sigma}g_{\mu\nu}\partial_\sigma f'\Big].
\eea
The trace of eq.~(\ref{field1}) is
\be
\label{trace}
f'({\cal R}){\cal R}-2f({\cal R})=\kappa T,
\ee
where $T=g^{\mu\nu} T_{\mu\nu}$.
This is actually an algebraic equation in ${\cal R}$ which can generically be solved to give ${\cal R}$ as a function of $T$.  $f\propto {\cal R}^2$ is an exception, which leads to a conformally invariant theory \cite{fer, sot1}. This exception, as well as choices of $f$ for which eq.~(\ref{trace}) has no root will not be considered further (in this case there are also no solutions of the full field equations \cite{fer}). Expressing ${\cal R}$ as a function of $T$ via eq.~(\ref{trace}) and using the result to eliminate the ${\cal R}$ dependence in the right-hand side of eq.~(\ref{gammagmn}) expresses the independent connection algebraically in terms of the metric and the matter fields. One can then proceed and eliminate the connection from the field equations. See, for example, Ref.~\cite{Sotiriou:2009xt} for more details and the final form of the field equations.

This establishes that the connection does not carry any dynamics for $f({\cal R})$ action as mentioned in the Introduction.

\section{More general actions}
\label{main}

We would now like to explore the dynamics of more general Palatini theories of gravity.  Our aim is to illustrate that for actions which contain generic higher order curvature invariants the independent connection cannot be algebraically eliminated (differently from the restricted $f({\cal R})$ case).
However, let us first point out that, as mentioned in the Introduction, in Ref.~\cite{Olmo:2009xy} the following class of actions was considered
\be
\label{olmoaction}
S=\frac{1}{l_p^2} \int dx^4  \sqrt{-g}f({\cal R}, {\cal R}_{\mu\nu}{\cal R}^{\mu\nu})+S_M(\psi,g_{\mu\nu}),
\ee
and there it was claimed that the connection can indeed be eliminated in such theories. This claim would obviously contradict our previous statement: even though action (\ref{olmoaction}) is restricted, it is still much more general than those in the $f({\cal R})$ class. In what follows we shall show that this contradiction is due to an implicit and unjustified assumption made in Ref.~\cite{Olmo:2009xy} regarding the symmetries of the Ricci tensor in Palatini theories.

We start by recalling that the Ricci tensor is given in term of the connection as
\bea\label{ricci}
{\cal R}_{\mu\nu}=\partial_{\lambda}\Gamma^{\lambda}_{\;\;\mu\nu}-\partial_{\nu}\Gamma^{\lambda}_{\;\;\mu\lambda}+\Gamma^{\lambda}_{\;\;\sigma\lambda}\Gamma^{\sigma}_{\;\;\mu\nu}-\Gamma^{\lambda}_{\;\;\sigma\nu}\Gamma^{\sigma}_{\;\;\mu\lambda}.
\eea
We can define the non-metricity of the connection as
\be
\label{nonm}
Q_{\lambda\alpha\beta}=-\nabla_{\lambda}g_{\alpha\beta}.
\ee

For what comes next we will restrict ourselves to a symmetric connection for simplicity. One could easily generalize our approach to include non-symmetric connections. However, it it obvious that if our claim is true for a symmetric connection it will continue to be true for a non-symmetric one.
A symmetric connection can be written as
\bea
\label{GQ}
\Gamma^{\rho}_{\phantom{a}\alpha\beta}=\big\{^{\rho}_{\phantom{a}\alpha\beta}\big\}+\frac{1}{2} g^{\rho\lambda}\left[Q_{\alpha\beta\lambda}+Q_{\beta\alpha\lambda}-Q_{\lambda\alpha\beta}\right],
\eea
where $\big\{^{\rho}_{\phantom{a}\alpha\beta}\big\}$ denotes the Levi-Civita connection of $g_{\mu\nu}$.
The non-metricity vector is defined as
\be
\label{nonmv}
Q_{\mu}=\frac{1}{4}Q_{\mu\nu}^{\phantom{ab}\nu}.
\ee
Then, for a symmetric connection, the antisymmetric part of the Ricci tensor is given by
\be
\label{antiricci}
\mathcal{R}_{[\alpha\beta]}=-\partial_{[\beta}\Gamma^{\lambda}_{\;\;\alpha]\lambda}=-2\nabla_{[\beta}Q_{\alpha]}
\ee
It should then be clear that ${\cal R}_{\mu\nu}$ is not necessarily symmetric even for a symmetric connection. Further restrictions on the non-metricity would have to be imposed to achieve that, which would restrict the connection.

Consider now the action
\be
\label{actionl4}
S=\frac{1}{l_p^2} \int dx^4  \sqrt{-g}\left[ {\cal R} +l_p^2 {\cal R}_{\mu\nu}(a {\cal R}^{\mu\nu}+b {\cal R}^{\nu\mu})\right]
\ee
Clearly this is not the most general action one could think of, but it is general enough for our purposes and simple enough to make the calculations tractable. Note that as long as ${\cal R}_{[\mu\nu]}\neq 0$ the last two terms are not equal. In fact (\ref{actionl4}) can be written as
\bea
\label{actionl42}
S=\frac{1}{l_p^2} \int dx^4 \sqrt{-g}\!\!\!\!\!&&\Big[{\cal R}+ c_1\, l_p^2\, {\cal R}_{(\mu\nu)}{\cal R}^{(\mu\nu)}\nn\\&&\qquad\qquad+  c_2\, l_p^2\, {\cal R}_{[\mu\nu]}{\cal R}^{[\mu\nu]}\Big],
\eea
where $c_1=a+b$ and $c_2=a-b$. Note also that for $b=0$, or $c_1=c_2$, action (\ref{actionl4}) reduces to the simplest model within the class given in action (\ref{olmoaction}), {\em i.e.}, to the case where $f$ is linear in both invariants.

We now vary the action independently with respect to the metric and the connection. The variation with respect to the metric yields
\bea
\label{geq2}
&&{\cal R}_{(\mu\nu)}-\frac{1}{2}{\cal R}g_{\mu\nu} +2c_1\, l_p^2 \, {\cal R}_{(\alpha\mu)}{\cal R}_{(\beta \nu)}g^{\alpha\beta} \nonumber \\&&\qquad+2 c_2\, l_p^2 \, {\cal R}_{[\alpha\mu]}{\cal R}_{[\beta \nu]}g^{\alpha\beta}-\frac{1}{2}c_1\,l_p^2 \, {\cal R}_{(\alpha\beta)}{\cal R}^{(\alpha\beta)}g_{\mu\nu}\nn\\&&\qquad\qquad-\frac{1}{2} c_2\,l_p^2\, {\cal R}_{[\alpha\beta]}{\cal R}^{[\alpha\beta]} g_{\mu\nu} =\kappa T_{\mu\nu}.
\eea
The variation with respect to the connection yields
\bea
\label{geq3}
&&-\nabla_\lambda\left[\sqrt{-g}\left(g^{\mu\nu}+2\,c_1\, l_p^2 {\cal R}^{(\mu\nu)}\right)\right]+\nabla_\sigma\left(\sqrt{-g}g^{\sigma(\mu}\right)\delta^{\nu)}_{\;\;\lambda}\nn\\
&&\quad+c_1\, l_p^2\, \nabla_{\sigma} \left[\sqrt{-g} {\cal R}^{(\mu\sigma)}\delta^\nu_{\phantom{a}\lambda}+ \sqrt{-g} {\cal R}^{(\nu\sigma)}\delta^\mu_{\phantom{a}\lambda}\right]\\&&\qquad +c_2\, l_p^2\, \nabla_{\sigma} \left[\sqrt{-g} {\cal R}^{[\mu\sigma]}\delta^\nu_{\phantom{a}\lambda}+\sqrt{-g} {\cal R}^{[\nu\sigma]}\delta^\mu_{\phantom{a}\lambda}\right]=0.\nn
\eea
Eq.~(\ref{geq3}) can be simplified by taking its trace and using it to replace the terms containing divergences. This leads to
\bea
\label{geq4}
&&\nabla_\lambda\left[\sqrt{-g}\left(g^{\mu\nu}+2\,c_1\, l_p^2 {\cal R}^{(\mu\nu)}\right)\right]\\&& +\frac{2}{3}c_2\, l_p^2 \,\nabla_{\sigma} \left[\sqrt{-g} {\cal R}^{[\mu\sigma]}\delta^\nu_{\phantom{a}\lambda}+\sqrt{-g} {\cal R}^{[\nu\sigma]}\delta^\mu_{\phantom{a}\lambda}\right]=0.\nn
\eea

Eqs.~(\ref{geq2}) and (\ref{geq4}) should reduce to eqs.~(3) and (4) of Ref.~\cite{Olmo:2009xy} for a linear function $f$ when we set $b=0$ or $c_1=c_2=a$. This is not the case however. The two sets of equations actually differ by terms including ${\cal R}_{[\mu\nu]}$. The fact that ${\cal R}_{[\mu\nu]}$ does not generically vanish for an independent connection, even a symmetric one as shown above, seems to have been overlooked in Ref.~\cite{Olmo:2009xy} and subsequently in Refs.~\cite{olmo2,new}. Hence, these terms were ignored there.\footnote{In Refs.~\cite{Allemandi:2004wn,Li:2007xw}, on the other hand, ${\cal R}_{\mu\nu}$ was explicitly assumed to be symmetric {\em a priori}.}

If one would indeed make the assumption that  ${\cal R}_{[\mu\nu]}=0$ then, for any values of $a$ and $b$ the system of equations would reduce to
\bea
\label{geq2sym}
&&{\cal R}_{(\mu\nu)}-\frac{1}{2} \left({\cal R}+c_1\, l_p^2\, {\cal R}_{(\alpha\beta)}{\cal R}^{(\alpha\beta)}\right) g_{\mu\nu}\nonumber \\&&\qquad\qquad+2 c_1\, l_p^2\, {\cal R}_{(\alpha\mu)}{\cal R}_{(\beta \nu)}g^{\alpha\beta}=\kappa T_{\mu\nu},\\
\label{geq4sym}
&&\nabla_\lambda\left[\sqrt{-g}\left(g^{\mu\nu}+2\,c_1\, l_p^2\, {\cal R}^{(\mu\nu)}\right)\right]=0.
\eea
The assumption that ${\cal R}_{[\mu\nu]}=0$ is equivalent to the requirement
\be
\label{condQ}
\nabla_{[\nu}Q_{\mu]}=0,
\ee
which essentially would mean that $Q_n$ is the gradient of a scalar.
Interestingly, one gets the exact same equations by assuming that $a=b$ or $c_2=0$ (which is different than the case considered in Ref.~\cite{Olmo:2009xy,olmo2,new}), without imposing any constraints on ${\cal R}_{[\mu\nu]}$ and consequently on the non-metricity. This choice of parameters correspond to an action which depends only on ${\cal R}_{(\mu\nu)}$.

Let us concentrate on these two cases for the moment, for which one can indeed apply the arguments of Ref.~\cite{Olmo:2009xy}. Notice that eq.~(\ref{geq2sym}) is actually an algebraic equation in ${\cal R}_{(\mu\nu)}$. That is to say, one could solve algebraically for the components of ${\cal R}_{(\mu\nu)}$, in terms of the components of $T_{\mu\nu}$ and $g_{\mu\nu}$ (even though it might not be possible to express the result in tensorial form). This could also be seen by thinking of eq.~(\ref{geq2sym}) as a matrix equation. Hence, ${\cal R}_{(\mu\nu)}$ in eq.~(\ref{geq4sym}) can be thought of as depending only on the matter fields and the metric, not on the connection.

Now, eq.~(\ref{geq4sym}) can be written as
\be
\label{conhmn}
\nabla_\lambda\left[\sqrt{-h}h^{\mu\nu}\right]=0,
\ee
where $h_{\mu\nu}$ is a symmetric metric implicitly defined via the relationship
\be
\label{hmn}
\sqrt{-h}h^{\mu\nu}=\sqrt{-g}\left(g^{\mu\nu}+2\,c_1\, l_p^2\, {\cal R}^{(\mu\nu)}\right).
\ee
Eq.~(\ref{conhmn}) implies that the independent connection is the Levi-Civita connection of $h_{\mu\nu}$. Since  $h_{\mu\nu}$ can be expressed in terms of the $g_{\mu\nu}$ and $T_{\mu\nu}$ one can then use the steps listed here in order to completely eliminate the independent connection $\Gamma^\lambda_{\phantom{a}\mu\nu}$.

As mentioned above, what was just described works for the specific choice of parameters $a=b$ or $c_2=0$ or if ones imposes {\em a priori} that ${\cal R}_{[\mu\nu]}=0$, which corresponds to  eq.~(\ref{condQ}). In the latter case, one would think that eq.~(\ref{condQ}) might impose an extra condition. However, it is trivially satisfied when eq.~(\ref{geq4sym}), or better yet eq.~(\ref{conhmn}) is satisfied. That is because a {\em sufficient} condition for a symmetric connection to lead to a symmetric Ricci tensor is for it to be the Levi-Civita connection of {\em some} metric. This can be easily shown by replacing the Levi-Civita expression for a connection in eq.~(\ref{antiricci}).

Even though we derived the results presented above using an action linear in Ricci squared invariants, there is no reason to believe that they are not more general than that. In fact, one should be able to eliminate a symmetric connection, in favor of the matter field and the metric, whenever only invariants constructed with the symmetric part of the Ricci tensor are considered in the action, {\em e.g.}~for Lagrangians of the form $f({\cal R},{\cal R}^{(\mu\nu)}{\cal R}_{(\mu\nu)})$. However, this is not the case for actions of the form $f({\cal R},{\cal R}^{\mu\nu}{\cal R}_{\mu\nu})$ as claimed in Ref.~\cite{Olmo:2009xy}. 

Let us see that in more detail. We return to more generic choices of parameters. Since the antisymmetric part of the Ricci enters the field equations now, the situation changes radically. Eq.~(\ref{geq2}) cannot be used to algebraically determine the full Ricci tensor, even at the component level, in term of the matter fields and the metric. Recall that if the Ricci is not assumed to be symmetric it has 16 independent components and eq.~(\ref{geq2}) corresponds to just 10 component equations. This is enough to argue that the presence of derivatives of ${\cal R}_{\mu\nu}$ in eq.~(\ref{geq3}) will make this equation a dynamical one in the independent connection. Therefore, one will not be able to eliminate the connection algebraically anymore.

As a simple but characteristic example let us consider the specific choice $a=-b$, or $c_1=0$, in which case the equations reduce to
\bea
\label{geq2anti}
\!\!\!{\cal R}_{(\mu\nu)}-\frac{1}{2} \Big({\cal R}&+& c_2\, l_p^2 \, {\cal R}_{[\alpha\beta]}{\cal R}^{[\alpha\beta]}\Big) g_{\mu\nu}\nonumber \\&+&2 c_2\,l_p^2 \, {\cal R}_{[\alpha\mu]}{\cal R}_{[\beta \nu]}g^{\alpha\beta}=\kappa T_{\mu\nu}, \\
\label{geq4anti}
\!\!\!\nabla_\lambda\left[\sqrt{-g}g^{\mu\nu}\right] &+&\frac{2}{3}c_2\, l_p^2 \,\nabla_{\sigma} \left[\sqrt{-g} {\cal R}^{[\mu\sigma]}\right]\delta^\nu_{\phantom{a}\lambda}\nn\\&+&\frac{2}{3}c_2\, l_p^2 \,\nabla_{\sigma} \left[\sqrt{-g} {\cal R}^{[\nu\sigma]}\right]\delta^\mu_{\phantom{a}\lambda}=0.
\eea
Contracting eq.~(\ref{geq4anti}) with the metric yields
\be
\label{traceanti}
g_{\mu\nu}\nabla_\lambda\left[\sqrt{-g}g^{\mu\nu}\right] =-\frac{4}{3}c_2\, l_p^2 \,g_{\lambda\nu} \nabla_{\mu} \left[\sqrt{-g} {\cal R}^{[\nu\mu]}\right].
\ee
On the other hand, one can straightforwardly show that
\be
\label{trick}
\nabla_{\mu} \left[\sqrt{-g} {\cal R}^{[\nu\mu]}\right]=\sqrt{-g}\bar{\nabla}_{\mu} \left[{\cal R}^{[\nu\mu]}\right],
\ee
where $\bar{\nabla}_\mu$ denote the covariant derivative defined with the Levi-Civita connection of $g_{\mu\nu}$.
Using eq.~(\ref{trick}) and eqs.~(\ref{nonm}) and (\ref{nonmv}), one can rewrite eq.~(\ref{traceanti}) as
\be
\label{vectoreq}
c_2\, l_p^2 \,\bar{\nabla}_{\mu} \left[{\cal R}^{[\nu\mu]}\right]-3Q^{\nu}=0,
\ee
while eq.~(\ref{geq4anti}) takes the simple form
\be
\label{geq4antiend}
Q_{\lambda\mu\nu}=2g_{\mu\nu} Q_\lambda-2g_{\lambda\mu} Q_\nu-2g_{\lambda\nu} Q_\mu.
\ee
Thus, the non-metricity can now be fully determined in terms of $Q_\nu$. The independent connection is then given by
\bea
\label{aside}
\Gamma^{\lambda}_{\phantom{a}\mu\nu}=\big\{^{\lambda}_{\phantom{a}\mu\nu}\big\}-3g_{\mu\nu}Q^{\lambda}+\delta^\lambda_{\phantom{a}\mu} Q_{\nu}+\delta^\lambda_{\phantom{a}\nu} Q_{\mu},
\eea
and ${\cal R}_{(\mu\nu)}$ can be expressed in terms of the Ricci tensor of $g_{\mu\nu}$, $R_{\mu\nu}$ and $Q_\nu$ as
\be
\label{mricci}
{\cal R}_{(\mu\nu)}=R_{\mu\nu}-3g_{\mu\nu} \bar{\nabla}_\sigma Q^\sigma-6 Q_\mu Q_\nu.
\ee
Taking a divergence of eq.~(\ref{vectoreq}) on can show that 
\be
\bar{\nabla}_\nu Q^\nu=0
\ee
Thus, eqs.~(\ref{geq2anti}) and (\ref{geq4anti}) are equivalent to the more familiar system
\bea
\label{geq2antieq}
&&R_{\mu\nu}-\frac{1}{2}Rg_{\mu\nu}=  -s\, \kappa \, F_{\alpha\mu}F_{\beta \nu}g^{\alpha\beta}+s\frac{1}{4} \kappa F_{\alpha\beta}F^{\alpha\beta} g_{\mu\nu}\nn\\&&\qquad\quad+\kappa\,m^2\,A_{\mu}A_{\nu}-\frac{1}{2} \kappa\,m^2\,A^{\sigma}A_\sigma g_{\mu\nu}+\kappa T_{\mu\nu}, \\
\label{geq4anti2eq}
 &&\bar{\nabla}_{\mu} F^{\mu\nu}+s\,m^2A^{\nu}=0.
\eea
where $F_{\mu\nu}=2\partial_{[\mu}A_{\nu]}$, $A_{\mu}=\sqrt{|c_2|/(4\pi)} \,Q_{\mu}$ and $m^2=3/(|c_2|l_p^2)$ and $s=sign(c_2)$. One can use these redefinitions and eqs.~(\ref{mricci}) and (\ref{antiricci}) to rewrite action (\ref{actionl42}) when $c_1=0$ as
\bea
\label{actionl4eq}
S=\frac{1}{l_p^2} \int dx^4 \sqrt{-g}R+S_F+S_M(\psi,g_{\mu\nu}),
\eea
where
\be
S_F=8\pi \int dx^4 \sqrt{-g}\Big[s\,\frac{1}{2}\, F_{\mu\nu}F^{\mu\nu}-m^2\,A^\mu A_\mu \Big].
\ee
One can easily verify that eqs.~(\ref{geq2antieq}) and (\ref{geq4anti2eq}) can be straightforwardly derived by varying action (\ref{actionl4eq}) with respect to $g_{\mu\nu}$ and $A_\mu$ respectively. Action (\ref{actionl4eq}), and consequently also action (\ref{actionl42}) with $c_1=0$, correspond to general relativity with matter and a massive vector field, also know as the Einstein--Proca field. This specific example was actually considered by Buchdahl in Ref.~\cite{buch}, where action (\ref{actionl4}) with $a=-b$ was proposed as a ``geometrization''   of the Einstein--Proca field.

One should have $s=-1$, {\em i.e.}~$c_2$ negative, for the vector field to not be a ghost and the theory to be quantum mechanically stable. This choice leads also to classical stability [our signature here is $(-+++)$].
In any case, irrespective of its physical relevance, this theory serves as a simple example of how higher order curvature invariants introduce extra degrees of freedom. It also demonstrates through the restriction in the sign of $c_2$ how the dynamics of these extra degrees of freedom can potentially lead to pathologies.  

As an aside, note that the connection given in eq.~(\ref{aside}) is a typical example of a symmetric connection for which ${\cal R}_{[\mu\nu]}\neq 0$: $A_\nu$ satisfies eq.~(\ref{geq4anti2eq}) which is well known to admit non-constant solutions. Because of the relation between $A_\nu$ and $Q_\nu$ and the relation between ${\cal R}_{[\mu\nu]}$ and $Q_\nu$ given in eq.~(\ref{antiricci}), one can easily infer that the theory admits solutions with ${\cal R}_{[\mu\nu]}\neq 0$.

\section{Conclusions}
\label{conc}

We have considered generalized Palatini theories of gravity, {\em i.e.},~theories with a connection which is independent of the metric and an action allowed to contain higher order curvature invariants than the Ricci scalar of this connection. We have shown that, unlike Palatini $f(R)$ theories, this connection does carry dynamics and cannot be algebraically eliminated. We gave as a simple, known, example the specific choice of action that is dynamically equivalent to the Einstein--Proca system (Einstein gravity plus a massive vector field). We also identified some specific actions which constitute exceptions, and for which the  independent connection can indeed be algebraically eliminated.

Our results disagree with those of Refs.~\cite{Olmo:2009xy,olmo2,new}. The reason appears to be that in Refs.~\cite{Olmo:2009xy,olmo2,new} the fact that the Ricci tensor of a symmetric connection is not necessarily symmetric unless extra constraint are imposed has been overlooked or it has been implicitly assumed that the Ricci tensor is indeed symmetric due to some restriction on the connection. 

We have not considered here theories where the independent connection is coupled to the matter as this will be the subject of a separate publication \cite{metaffdyn}.\\

\begin{acknowledgments}
We would like to thank Gonzalo Olmo for discussions. T.P.S. was supported in part by STFC and in part by a Marie Curie International Incoming Fellowship.
\end{acknowledgments}


\end{document}